\documentclass[10pt,twocolumn,english]{article}
\usepackage{mathptmx}
\usepackage[T1]{fontenc}
\usepackage[latin9]{inputenc}
\usepackage[paperwidth=24.51cm,paperheight=34.51cm]{geometry}
\usepackage{float}
\usepackage{calc}
\usepackage{textcomp}
\usepackage{amsmath}
\usepackage{graphicx}
\usepackage[numbers]{natbib}

\makeatletter

\newcommand{\lyxmathsym}[1]{\ifmmode\begingroup\def\b@ld{bold}
  \text{\ifx\math@version\b@ld\bfseries\fi#1}\endgroup\else#1\fi}

\providecommand{\tabularnewline}{\\}

\AtBeginDocument{
  
}

\makeatother

\usepackage{babel}
\begin{document}
\pagestyle{empty}

\title{An analytic study of a distributed EDCA-based QoS mapping for layered
video delivery in WLAN}

\author{Lamia Romdhani and Amr Mohamed\\
Computer Science and Engineering Department, Qatar University, Doha,
Qatar\\
\{lamia.romdhani,amrm\}@qu.edu.qa \\
}

\date{~}

\maketitle
\thispagestyle{empty}

\begin{center}
\textbf{\large Abstract}
\par\end{center}{\large \par}

\emph{\,One of the key challenges in multimedia networks is video
delivery over wireless channels. MRC (Multi-Resolution Coding) Layered
video, divides video into a base layer and multiple enhancement layers.
In this paper, we aim to improve video quality, impacted by high channel
contention, through mapping individual video layers to EDCA (Enhanced
Distributed Channel Access) access categories in order to maximize
the average number of reconstructed video layers. We propose an adaptive
cross layer video layers mapping technique that optimally enhances
the QoS of wireless video transmission over IEEE 802.11e EDCA priority
queues. The optimization is based on a dynamic program that takes
into account the EDCA parameters and the layered dependency nature
of layered video delivery. Our proposed technique makes use of a channel
delay estimation model and an estimation of average video useful layers
delivered. The optimal mapping strategies are selected by an optimization
module based on the information from the analytical model. The accuracy
of our optimized mapping technique performance is verified through
extensive simulations. The obtained results illustrate significant
trade-off between complexity and delivered video quality for canonical
mapping schemes.}


\section{\label{sec:Introduction}Introduction}

The requirement of high video delivery quality over wireless networks
is increasing day-by-day. Layered video, scalable video, and multiple
resolutions coding (MRC), all refer to encoding techniques that fragment
a video stream into a base layer and enhancement layers \citep{layered-video}.
The base layer is necessary for decoding the video stream, whereas
the enhancement layers improve its quality. This approach is useful
for wired multicast, where a receiver with a congested link can download
only the base layer, and avoid packets from other layers. With wireless,
all layers share the medium. Thus, the enhancement layers reduce the
bandwidth available to the base layer and further reduce the performance
of poor receivers. The Quality-of-Service (QoS) of 802.11e \citep{80211e_standard}
is achieved by providing different classes of frames with different
priorities when accessing the radio channel. In the basic EDCA scheme,
the video traffic is mapped automatically to two access classes. In
this paper, we describe a distributed and adaptive cross-layer dynamic
mapping techniques that map the arriving video packets into different
EDCA Access Categories (ACs) to optimize layered video delivery by
maximizing the expectation of the number of video layers received..

The remainder of this paper is organized as follows. We devote Section
\ref{sec:related-works}, for reviewing some related works from the
literature for enhancing video delivery in wireless networks. Our
proposal will be described in Section 3. We provide a deeper analysis
of the main obtained results in Section 4. Section 5 summarizes the
paper and outlines the future works.

\section{\label{sec:related-works}Related Works}

Many works have been presented in the literature to enhance video
delivery in wireless environment. They described many features based
on rate allocation, channel quality estimation, retry limit adaptation,
queue length estimation..etc. In \citep{key-6} the cross-layer QoS-optimized
EDCA adaptation algorithms take into account the unequal error protection
characteristics of video streaming, the IEEE 802.11e EDCA parameters
and the lossy wireless nature. It makes use of two models, video distortion
model and channel throughput estimation model to predict the video
quality. The convex nature of optimization problem remains an open
research issue. The work of rate allocation becomes challenging since
heterogeneity exists in both the rate utilities of video streams and
in wireless link qualities. In distributed manner the task can be
performed, as many times the system lack centralized control. In \citep{rate-allocation},
an optimization framework to distribute video rate allocation over
wireless is proposed, taking into account this challenge. In \citep{key-13}
the authors investigate the packet loss behavior in the IEEE 802.11e
wireless local area networks (WLANs) under various retry limit settings.
Considering scalable video traffic delivery over the IEEE 802.11e
WLANs, the presented study shows the importance of adaptiveness in
retry limit settings for the Unequal Loss Protection (ULP) design.
Based on the study, they present a simple yet effective retry limit
based ULP which adaptively adjusts the retry limit setting of the
IEEE 802.11e medium access control protocol to maintain a strong loss
protection for critical video traffic transmission. 

A new packet scheduler in cross layer environment for GSM/EDGE systems
to improve QoS support of multiclast data services is proposed in
\citep{key-4}. The algorithm minimizes a prescribed cost functions
given the current channel qualities and delay states of the packets
in the queue. A cross-layer optimization for video streaming over
wireless multimedia sensor networks is attempted in \citep{cross-layer-video}.
In 802.11s mesh networks, packets are differentiated and higher priorities
are given to forward packets. When queue length of AC2 fills up, forward
packets are remapped to lower access category AC3.

Weighted Fair Queuing \citep{key-8} is efficient for wireless channels.
It assigns weight for different flows and calculates the departure
time based on the weights. Assigning weight to the individual flows
helps in prioritizing the video packets and sending the packets in
the flow which has more weight. 

Forward Error Correction \citep{key-9} is used to reduce the number
of packets lost. This is done by adding redundant number of packets
to the video sequences. The challenge is to add optimum number of
packets suitable for both channel availability and queue length. An
adaptive video packet scheduling algorithm used in WLAN is proposed
in \citep{key-10}. The data transmitted over the wireless channel
should be reduced as much as we can consider of the limitation of
wireless bandwidth, but not the video-quality. If the network load
becomes higher and higher, the access point must compare the multiple
video streams and find which one should be transmitted first. Unlike
previous works, this paper addresses a simple and adaptive distributed
mapping strategy based on EDCA access scheme. We describe an analytical
model for selecting the best strategy to map video layers to each
AC in order to maximize the reconstructed video layers taking into
account the wireless channel contention model and the video layer
dependency.

\section{\label{sec:system-model}System Model}

In this Section we describe in details our analytical model for wireless
layered video delivery. Our solution is fully distributed as in \citep{MRC_distributed_EdCA}.
It is based on EDCA mechanism, witch provides a differentiated, distributed
access to the medium using different priorities for different types
of traffic \citep{80211e_standard}. We consider a layered video source
encoded into a base layer which contains the most important information,
and enhanced layers that provide additional information for better
video quality. We assume that the video layers have the same constant
bitrate. Our aim is to select, from exhaustive search results, the
best mapping strategy of video layers to different EDCA ACs, which
decreases the dropping probability and improves the expected number
of useful layers delivered to the destination, regarding different
settings of EDCA parameters and traffic load.

Basically, an EDCA channel access function uses Arbitration Inter-frame
Space $(AIFS[AC])$, Contention Window with its minimum and maximum
value $CW_{min}[AC]$ and $CW_{max}[AC]$ respectively instead of
DIFS,$CW_{min}$ and$CW_{max}$, of the DCF (Distributed Coordination
Function) respectively, for the contention process to transmit a packet
that belongs to $AC$. These parameters can be used in order to differentiate
the channel access among different priority traffic. The channel access
priority goes from $AC_{4}$, $AC_{3}$, $AC_{2}$, up to the highest
priority $AC_{1}$. As the priority of the $AC$ increases, the values
of the MAC parameters become smaller. Thus the $AC$ with the shorter
contention period has more priority to occupy the channel.

\subsection{Analytical study}

For ease of understanding, we present a simple EDCA model under saturation
condition \citep{model-edca}. This model estimates the following:
1) interface queue dropping probability that computes the packets
drop due to queue overflow, 2) a delay model that accounts for all
events that contribute to the access delay, and finally 3) we derive
the expected number of useful layers successfully delivered to the
destination node regarding a defined layered video mapping strategy.
These parameters capture the influence of the $CWmin[AC_{i}]$ and
$CWmax[AC_{i}]$, $AIFS$, and Transmission Opportunity $(TXOP)$
mechanisms. Moreover, we define the concepts of mapping, ordered mapping,
exhaustive mapping, and canonical mapping. We compute the complexity
of each mapping concept. Then, we present how the best mapping strategy
is defined. Furthermore, we discuss the tradeoff between complexity
and performance enhancement of layered video delivery over wireless
network.

We assume that we have $N$ video users (or subscriber stations SS),
and $n_{AC_{i}}$ number of $AC_{i}$ contending for the channel access.
$RL_{i,retry}$ is the maximum retry limits of access category $AC_{i}$.

\subsubsection{EDCA model}

The proposed model is based on the Markov chain introduced in \citep{bianchi,model-edca}.
It extends the probability formulas to support differential $TXOP_{Limit}$
parameter in the different computed performance metrics \citep{eda-txop}.
In the following, we denote by $\tau_{i}$ the probability that a
node in the $AC_{i}$ transmits during a generic slot time and by
$p_{i}$ the probability that $AC_{i}$ senses the medium busy around
it. The $\tau_{i}$ takes into account both internal and external
collision.

\begin{eqnarray}
\tau_{i} & = & \left(\sum_{j=0}^{RL_{i,retry}}b_{i,j,0}\right)*\prod_{h<i}\left(1-\tau_{h}\right)\nonumber \\
 & = & \left(b_{i,0,0}*\frac{1-p_{i}^{RL_{i,retry+1}}}{1-p_{i}}\right)\label{eq:prob_transmission}
\end{eqnarray}

\begin{equation}
b_{i,0,0}=\frac{1}{\sum_{j=0}^{RL_{i,retry}}\left[1+\frac{1}{1-p_{i}}\sum_{k=1}^{W_{i,j}-1}\frac{W_{i,j}-k}{W_{i,j}}\right]p_{i}^{j}}\label{eq:b00}
\end{equation}

Where $b_{i,0,0}$ is the initial state of the $AC_{i}$. We follow
the basic EDCA backoff increase scheme \citep{bianchi}. From the
point of view of one wireless node, the probability $\tau$ that the
node access to the medium is:

\begin{equation}
\tau=1-\left(\prod_{i=1}^{4}\left(1-\tau_{i}\right)\right)\label{eq:pb_sta_trx}
\end{equation}

We aim to derive for a given $AC_{i}$ , the formulas of saturation
throughput, delay, and queue dropping probability. We focus here on
packet dropping due to both queue overflow, and reaching the maximum
retry limit. We assume that the frame corruptions are only due to
collisions, thus no channel error is considered.

The collision probability due to both internal and external collisions
is, defined as follows, for an $AC_{i}$ :

\begin{equation}
p_{coll,i}=1-\left(1-\tau\right)^{N-1}\prod_{h\prec i}\left(1-\tau_{h}\right)\label{eq:pb_coll_Aci}
\end{equation}

Where $h\prec i$ means that $AC_{h}$ has higher priority than $AC_{i}$.

Let $p_{succ,i}$ be the probability that an $AC_{i}$ succeeds to
transmit a packet and $p_{succ}$ the probability that a node achieves
a successful transmission.

\begin{equation}
p_{succ,i}=N*\tau_{i}\left(1-p_{coll,i}\right)\label{eq:p_i_succ}
\end{equation}

We can obtain the total saturation throughput for the system as follows:

\begin{equation}
S_{i}=\frac{E\left[P_{i}\right]}{E\left[L\right]}\label{eq:throuput-1}
\end{equation}

Where $E[P_{i}]$ is the payload transmitted in a transmission period
for a class $i$, and $E\left[L\right]$ is the length of a transmission
period. According to \citep{bianchi,eda-txop} the throughput can
be defined as: 

\begin{equation}
S_{i}=\frac{p_{succ,i}\left(K_{TXOP_{i}}+1\right)E\left[length_{data}\right]}{\left(1-p_{busy}\right)\theta+p_{succ}T_{S_{i}}+p_{coll}T_{C}}\label{eq:throuput_saturation}
\end{equation}

Where $E\left[length_{data}\right]$is the average data packet length,
$p_{succ}$ the probability that a station transmit successfully,
$p_{coll}$ the probability that a collision occurs foe station ,
$T_{S_{i}}$ the transmission time, $T_{C}$ the collision time, $K_{TXOP_{i}}$
the number of packets transmitted during transmission opportunity
period, and $\theta$is the duration of the slot time.

The access delay for each $AC_{i}$ is defined as:

\begin{equation}
E\left[D_{i}\right]=\frac{E\left[P_{i}\right]}{S_{i}}\label{eq:delay}
\end{equation}

\begin{itemize}
\item Frame-dropping probability analysis
\end{itemize}
Let $P_{i,drop}$ be the probability of packet drops (see Eq.\ref{eq:prob_dropp})
\begin{eqnarray}
P_{i,drop}= & 1-(\left(1-P{}_{i,drop,coll}\right)*\nonumber \\
\label{eq:prob_dropp}\\
 & \left(1-p_{queue_{drop,i}}\right))\nonumber 
\end{eqnarray}

\begin{equation}
P{}_{i,drop,coll}=p_{i}^{L_{i,retry}+1}\label{eq:prob_dropp_hi}
\end{equation}

Where $P_{queue_{drop,i}}$ is the probability that a packet is dropped
due to the queue overflow, and $P{}_{i,drop,coll}$ represents the
probability of frame drops due to maximum retry limit \citep{model-edca}.
Let $K$ be the maximum size of the queue, and $\lambda_{i}$ is the
application rate of an $AC_{i}$. We assume an exponential arrivals
and departures of packets in the queue. So the service rate is $\mu=\frac{1}{E\left[D_{i}\right]}$
and the traffic intensity or the offered load is defined as $\rho:$

\begin{equation}
\rho_{i}=\frac{\lambda_{i}}{\mu_{i}}\label{eq:rho}
\end{equation}

We consider the M/G/1/K state transition diagram. Thus, $P_{queue_{drop,i}}$
is the probability that there are $K$ packets in the queue at an
arbitrary time:
\begin{equation}
P_{queue_{drop,i}}=\frac{\rho_{i}^{K}\left(1-\rho_{i}\right)}{1-\rho_{i}^{K+1}}\label{eq:dropQue}
\end{equation}

\subsection{EDCA-based layered video delivery model }

We first define the following concepts:

\textbf{Layered video concept:} In video coding schemes such as H.264/AVC,
the video content is partitioned into sequences of pictures, referred
to as groups of pictures (GOPs), each beginning with an independently
decodable intra-coded picture. A typical duration for a GOP is 1-2
seconds. Each GOP contains many pictures or frames. A GOP is divided
into a sequence of packets for delivery over the network. Although
a single frame may span multiple packets, or a single packet may contain
more than one frame, we can assume that there will be multiple packets
for a GOP, and in the case of constant bitrate video coding, the number
of packets per GOP will be constant throughout a sequence. Layered
video concept as MRC (Multi-Resolution Coding), divides the video
into a base layer and multiple enhancement layers. The base layer
can be decoded to provide a basic quality of video while the enhancement
layers are used to refine the quality of the video. If the base-layer
is corrupted, the enhancement layers become useless, even if they
are received perfectly. Moreover, in MRC, receiving the $K^{th}$layer
is only helpful if the previous $K\lyxmathsym{\textminus}1$ layers
have been received. Thus, in layered coding, the video content is
partitioned into multiple layers of sub-streams, and hence each GOP
can be thought of as consisting of several sequences of packets, one
for each layer. We assume that these sub-streams have a constant bitrate
\citep{MRC-constantrate}.

\subsubsection{Calculation of expected number of useful layers}

We aim to address an efficent video transmission scheme based on EDCA
medium access mechanism, that maximizes the number of useful layers
received at the destination node. We define the estimated number of
video layers based on the probabilities of the individual dropping
probability of each layer:

\begin{equation}
E[UL](L)=\sum_{r=1}^{L}r*\prod_{i=1}^{r}\left(1-P_{l_{i}}\right)*\prod_{h=r}^{L}P_{l_{h}}\label{eq:avgULGenr}
\end{equation}

Where $P_{l_{i}}$is the dropping probability of layer $l_{i}$ and
$L$ is the total number of video layers represented by\textbf{ $S(L)=\left\{ l_{1},l_{2},..l_{L}\right\} $
}. $P_{l_{i}}$depends on which $AC$ the layer $l_{i}$ is mapped
to. Thus, for all layers assigned to $AC_{1}$, the $P_{l_{i}}$ is
equal to $P_{1,drop}$, the layers assigned to $AC_{2}$, the $P_{l_{i}}$
is equal to $P_{2,drop}$, these mapped to $AC_{3}$, the $P_{l_{i}}$
$P_{3,drop}$ and these mapped to $AC_{4}$, the $P_{l_{i}}$ is equal
to $P_{4,drop}$. It can be shown that to calculate $E[UL](L)$, we
need two nested loops to calculate the product and the summation,
which deems the complexity of calculating E{[}UL{]} is $O(L^{2})$.
The $P_{l_{i}}$are pre-computed for all $r=1,....L$ regarding ACs
packet drop probabilities.

Having obtained the medium contention, the dropping probabilities,
and the packet collision probabilities from the model described in
the previous subsection, we can compute the expected number of useful
layers received, under different traffic load, for each mapping vector.
Let $n_{max}$ be the maximum number of $ACs$:$\left\{ AC_{1},AC_{2},..AC_{n_{max}}\right\} $
(for EDCA, $n_{max}=4$). We aim to map $S(L)$ to different set of
$ACs$. We define\textbf{ $M(L,n)=\{m_{AC_{1}},m_{AC_{2}},m_{AC_{3}},..m_{AC_{n}}\}$
}an arbitrary mapping vector, that maps $L$ layers to $n$ ACs $(1\leq n\leq n_{max})$.
Where $m_{AC_{i}}$ is the number of video layers selected from $S(L)$
and assigned to $AC_{i}$. This leads to $L=\sum_{i=1}^{n}m_{AC_{i}}$.
We aim to investigate the video performance of different mapping strategies
of video layers, within each GOP, to different EDCA $ACs$. We calculate
the expected number of video layers metric for each mapping vector.
Then, we select the best mapping strategy vector $M(L,n)$ regarding
the maximum estimated value of average useful layers. We believe that
considering this metric in our mechanism gives an accurate information
about video delivery quality. Furtheremore, we have to perform an
exhaustive search algorithm for all possible mapping strategies of
video layers to different EDCA $ACs$. The Complexity of search for
exhaustive mapping strategies $C_{exh_{mapp}}$ , when considering
four $ACs$ and $L$ layers, is:

\begin{eqnarray}
C_{exh_{mapp}}= & \sum_{i=1}^{4}\left(_{_{L-i}}{}^{L-1}\right)\nonumber \\
\label{eq:complexity_mapping}\\
= & 1+(L-2)+\frac{(L-1)(L-2)}{2}\nonumber \\
+ & \frac{(L-1)(L-2)(L-3)}{6}\nonumber 
\end{eqnarray}

Thus, from Equation \ref{eq:complexity_mapping}, we deduce that the
best strategy selection algorithm over exhaustive mapping has a high
complexity, which is about $C_{exh_{mapp}}=O(L^{3})$. When considering
$n$ $ACs$, $C_{exh_{mapp}}=O(L^{n-1})$. Hereafter, we aim to minimize
this complexity by extracting the group to which the best strategy
belongs. We divide the exhaustive mapping strategies into two groups:
canonical and non-canonical.

\textbf{Exhaustive mapping: }The exhaustive mapping defines all possibilities
of mapping vectors $\Delta(M)=\left\{ M(L,n):1\leq n\leq n_{max}\right\} $.

\textbf{Canonical mapping: }In the canonical mapping, the number of
layers assigned to each AC increases with the class priority level.
This leads to $m_{AC_{n}}\ge m_{AC_{n+1}}$ for any $n$: $1\leq n\leq n_{max}$
and $\{m_{AC_{n},}m_{AC_{n+1}}\}\subset M_{cano}(L,n)$. Where $M_{cano}(L,n)$
is a canonical mapping vector.

\textbf{Ordered mapping: }In the ordered mapping concept, if a video
layer $l_{i}$ is assigned to an $AC_{j}$, the layer $l_{i'}$, where
$i'>i$ should be assigned to $AC_{j'}$ where $j'\geq j$ . Recall
that, for simplicity, in our EDCA model we consider $AC_{j}$ has
higher priority than $AC_{j+1}$.

\textbf{Non-Canonical mapping: }The non-canonical vectors is: $\Delta(M)-M_{cano}(L,n)$.

\textbf{\textit{Lemma:}}

Considering our distributed environment and constant layers bitrate,
the optimal mapping vector exists in the canonical ordered mapping.

\textbf{\textit{Proof:}}

We proceed to prove the lemma by contradiction. Let's consider $M^{*}(L,n)=\{m_{AC_{1}},m_{AC_{2}},..m_{AC_{i}},..m_{AC_{j}}..m_{AC_{n}}\}$
the optimal mapping vector that gives the best $E[UL](L)$ where $m{}_{AC_{j}}>m{}_{AC_{i}}$.
Thus, $M^{*}(L,n)$ is a non-canonical mapping vector. Let $P_{drop}^{*}=P_{drop,m_{AC_{i}}}*P_{drop,m_{AC_{j}}}$
the dropping probability of layers assigned to $AC_{i}$ and $AC_{j}$.
Let $M^{'}(L,n)=\{m_{AC_{1}},m_{AC_{2}},..m_{AC_{i}}^{'},..m_{AC_{j}}^{'}..m_{AC_{n}}\}$
a mapping vector where $m_{AC_{i}}^{'}=m_{AC_{j}}$ and $m_{AC_{j}}^{'}=m_{AC_{i}}$$\Rightarrow$
$m^{'}{}_{AC_{i}}>m^{'}{}_{AC_{j}}$ . The dropping probability on
the layers assigned to $AC_{i}$ and $AC_{j}$ according to the vector
$M^{'}(L,n)$ is $P_{drop}^{'}=P_{drop,m_{AC_{i}}^{'}}^{'}*P_{drop,m_{AC_{j}}^{'}}^{'}$.
Regarding EDCA service differentiation medium access, $P_{i,drop}<P_{j,drop}$
(see Equation \ref{eq:prob_dropp}). Thus, when assigning more layers
to $AC_{j}$, more layers will be dropped than assigning the same
number of layers to $AC_{i}$. We obtain $P_{drop}^{'}<P_{drop}^{*}$
and therefore $E^{'}[UL](L)>E^{*}[UL](L)$ , which gives $M^{*}(L,n)$
is not the optimal mapping vector, and the optimal value exists for
$M^{\sim}(L,n)$ with $m{}_{AC_{i}}\geq m{}_{AC_{j}}$ $\surd$$1\leq i<j\leq n$. 

\textbf{Ordered mapping: }For a given mapping strategy, we suppose
that it exists $l_{i}$, witch is assigned to $AC_{n}$ and $l_{j}$,
$(j>i)$ is assigned to $AC_{m}$ has higher priority than $AC_{n}$).
We know that: $P_{succ,m}>P_{succ,n}$ (regarding the service differentiation
addressed with EDCA model), thus the probability that $l_{i}$ collides
is higher than the probability that $l_{j}$ collides, and so $l_{j}$
becomes useless even it is transmitted successfully when $l_{i}$
is lost. This leads to a decreasing on the number of average useful
video layers delivered. Thus, we have to ensure ordered layers mapping
to different $ACs$ to enhance video delivery quality.

Hereafter, we present the Optimal Canonical Ordered Mapping (OCOM)
algorithm that will calculate the Optimal mapping based on ordered
canonical mapping. Let $j$ the number of active $ACs$ considered
in the mapping vector $M(L,j)$, and $m_{i,j}$ is the maximum number
of layers assigned to $AC_{i}$, $\surd i$ ,$i\leq j$. Thus, $m_{i,j}$
is calculated as:

\begin{eqnarray}
m_{i,j}=\begin{cases}
\begin{array}{cccccc}
0 &  &  &  & j<i\\
m_{i-1,,j} &  & L\, mod\, j\leq i &  & j\geq i\\
m_{i-1,,j}-1 &  & L\, mod\, j>i &  & j\geq i
\end{array}\end{cases}\label{eq:max_m-1}
\end{eqnarray}

Where the mod function calculates the remainder of dividing $L$ by
$j$. Thus for one $AC$ , we obtain $m_{1,1}=L$ . Hereafter, we
describe the OCOM algorithm. For two $ACs$, we define:

$m_{22}=\begin{cases}
\begin{array}{ccc}
\frac{L}{2} &  & L\, mod\,2=0\\
\frac{L}{2}-1 &  & L\, mod\,2>0
\end{array}\end{cases}$ 

We present the OCOM algorithm description for $L$ layers and $n$
$ACs$ in Table I.

\textbf{Complexity:} Let $C_{E[UL]}$ the complexity of computing
$E[UL]$, $C_{OCOM}$ the complexity of OCOM algorithm, and $C_{L,n}$
the complexity of selecting the best mapping vector based on the maximum
$E[UL]$ having $L$ video layers and $n$ $ACs$. When considering
$4$ $ACs$ $C_{OCOM}=1+\frac{L}{2}+(1+2+..\frac{L}{3})+(1+2+..\frac{L}{4})$.
$C_{OCOM}=O(4L^{2})$. For $n$ $ACs$, $C_{OCOM}=O(nL^{2})$. We
calculte $C_{L,n}=C_{E[UL]}*C_{OCOM}$. Thus $C_{L,n}=O(nL^{4})$.

To optimize the computation time and to reduce the complexity of selecting
the best mapping strategy vector algorithm presented above, we define
a new dynamic program that can be used to calculate $E[UL]$ recursively
regarding the number of ACs used in the mapping strategy:

\begin{enumerate}

\item \textbf{For} $n=1$: all $L$ video layers are mapped to the
highest AC:

\begin{equation}
E_{dp}[UL]_{AC(1)}(l_{AC_{1}})=\sum_{i=1}^{l_{AC_{1}}}i*\left(1-P_{1,drop}\right)^{i}*P_{1,drop}^{l_{AC_{1}}-i}\label{eq:avgUL1}
\end{equation}

Where, $l_{AC_{i}}$is the number of video layers mapped to $AC_{i}$.
The complexity $C_{E_{dp}[UL]}$ of equation \ref{eq:avgUL1} is $O(l_{AC_{1}})$.
When considering one $AC$: $l_{AC_{1}}=L$, thus $C_{E_{dp}[UL]}=O(L)$.

\item \textbf{For} $n>1$: $(n$ $ACs)$: map the $L$ video layers
to $n$ $ACs$$:$$l_{AC_{1}}$ layers to $AC_{1}$and $($$l_{AC1}$
layers to $AC_{2})$ that we note by $AC_{1,2}$:

\begin{eqnarray}
E_{dp}[UL]_{ACs(n)}(l_{AC_{1,2,..,n}})=E_{dp}[UL]_{ACs(n-1)}(l_{AC_{1,2,..n-1}})\nonumber \\
*P_{n,drop}^{l_{AC_{n}}}+\prod_{h=1}^{n-1}\left(1-P_{h,drop}\right)^{l_{AC_{h}}}\label{eq:avgUL4ACs}\\
*(\sum_{i=1}^{l_{AC_{n}}}(i+\sum_{j=1}^{n-1}l_{AC_{j}})*\left(1-P_{n,drop}\right)^{i}*P_{n,drop}^{l_{AC_{n}}-i})\nonumber 
\end{eqnarray}

\end{enumerate}

In this case $C_{E_{dp}[UL]}=O(\sum_{j=1}^{n}l_{AC_{j}})$. $\sum_{j=1}^{n}l_{AC_{j}}=L$.
Therefore, the complexity obtained using the new recursive Equation
(\ref{eq:avgUL4ACs}) is equal to $O(L)$ regardless of the number
of $ACs$ considered in the mapping strategy. This complexity is lower
than the complexity obtained with Equation (\ref{eq:avgULGenr}).
We compute $E_{dp}[UL]$ for each canonical mapping possibility to
select the best mapping. Let $C_{L,n}^{'}$ the compexity of the best
strategy using Equation (\ref{eq:avgUL4ACs}). $C_{L,n}^{'}=O(nL^{3})$.

\begin{figure}[htbp]
\begin{centering}
\includegraphics[width=6cm,height=4.5cm,angle=360]{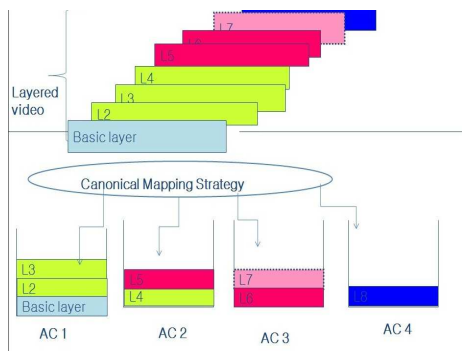}
\par\end{centering}

\caption{\label{cap:Adaptive-mapping}Canonical mapping EDCA-based architecture}
\end{figure}

An example of our EDCA-based model architecture, based on ordered
canonical mapping, is shown in Figure \ref{cap:Adaptive-mapping}.

Although the algorithm OCOM is practically feasible, we are seeking
to reduce the complexity to make it more practical and scalable to
the number of layers. Hence, we're proposing an optimal distributed
adaptive algorithm selecting a specific mapping strategy regarding
obtained results and analysis. This algorithm aims to maximize the
average useful layers delivered. We aim to obtain a good performance
close to the best canonical solution results, with low complexity.

\framebox{\parbox[c][15cm]{7cm}{%
\textit{Table I: }\textbf{\textit{OCOM algorithm:}}

$\,$

($E_{min}(L,n)$,$x_{min}$ )=OCOM(L,n)

$\,$$x_{min}=zeros(n)$

$\, E_{min}(L,n)=\infty$

For $i=1$ to $n$ 

($E_{min}(L,n)$,$x_{min}^{'}$ )= Call Map ($L$, $i$, $E_{min}(L,n)$,
$x_{min}$)

If $x_{min^{'}}(i)$==0 stop

\_\_\_\_\_\_\_\_\_\_\_\_\_\_\_\_\_\_

($E_{min}(L,n)$,$x_{min}^{'}$ )=Map (L, 1, $E_{min}(L,n)$, $x_{min}$
) 

$x_{min}^{'}$$=L$ 

caculate $E_{min}(L,1)$ 

\_\_\_\_\_\_\_\_\_\_\_\_\_\_\_\_\_\_

($E_{min}(L,n)$,$x_{min}^{'}$ )$=Map(L,2,E_{min}(L,n),x_{min})$\\
$x_{min}^{'}$$={x_{min},0}$ \\
Calculate $m_{2,2}$// using Equation (\ref{eq:max_m-1}).

For $i2=1:m_{2,2}$ 

$\,\,\,\,\,\,$Calculate $E(UL)(L)$ 

$\,\,\,\,\,\,$If $E<E_{min}$ 

$\,\,\,\,\,\,$$E_{min}=E$ 

$\,\,\,\,\,\, x_{min}^{'}=x$

\ldots{}\ldots{}\ldots{}\ldots{}\ldots{}\ldots{}\ldots{}\ldots{}\ldots{}\ldots{}\ldots{}\ldots{}\ldots{}\ldots{}\ldots{}\ldots{}

($E_{min}(L,n)$ ,$x_{min}^{'}$ )=Map$(L$, $k$, $E_{min}(L,n)$
, $x_{min})$ 

$x_{min}^{'}$= \{ $x_{min}$,0\} 

Calculate $m_{2,k}$ // using Equation (\ref{eq:max_m-1})

For $i2=1:m_{2,k}$

$\,\,\,\,\,\,$Calculate $m_{3,k}$ // using Equation (\ref{eq:max_m-1})

For $i3=1:m_{3,k}$ 

\ldots{}\ldots{}\ldots{}\ldots{}\ldots{}\ldots{}\ldots{}\ldots{}\ldots{}\ldots{}\ldots{}.

\ldots{}\ldots{}\ldots{}\ldots{}\ldots{}\ldots{}\ldots{}\ldots{}\ldots{}\ldots{}\ldots{}.. 

$\,\,\,\,\,\,$Calculate $m_{k,k}$ // using Equation (\ref{eq:max_m-1})

$\,\,\,\,\,\,$For $ik=1:m_{k,k}$ 

$\,\,\,\,\,$Calculate $E(UL)(L)$

$\,\,\,\,\,\,$If $E<E_{min}$ 

$\,\,\,\,\,\, E_{min}=E$

$\,\,\,\,\,\, x_{min}^{'}=x$%
}}

\textbf{Proposed algorithm description:}

\framebox{\begin{minipage}[t]{0.35\paperwidth}%
\begin{itemize}
\item step 1: Determine IFQ dropping probability and collision probability
regarding EDCA parameters and channel contention feedback.
\item step 2: Select the mapping strategy, regarding minimum tolerated packet
drop percentage threshold $\delta$ for each ACs: The number of layers
assigned to the queue cannot cause more than $\delta$ (arbitrary
parameter) of interface queue packet drops. We still assign layers
to the AC till the threshold is acheived. Then, we move to assign
the remaining layer to next ACs.
\item step 3: Calculate the average useful layer according to equations
(26)(27) for the selected canonical mapping.
\item step 4: The different metrics are periodically updated regarding real-time
channel varying conditions.\end{itemize}
\end{minipage}}

To perform step 2, that is described in the above sub-optimal algorithm,
we can calculate the number of layers $l_{i}$ to be assigned to $AC_{i}$
and then, the mapping vector such that the estimated dropping probability
is less then or equal to $\delta$. Hence, we compute first $\rho$
from Equation (\ref{eq:dropQue}) by fixing the threshold $\delta$
and so the IFQ dropping probability. Then, we can deduce $l_{i}$
from Equation (\ref{eq:rho}).

As wireless channel is time-varying, the dropping probabilities are
computed periodically based on the available wireless resources. The
mapping strategy is updated when the drops increase and the average
useful layer decreases.

\section{\label{sec:Model-validation}Model Validation}

In this section, we report different analysis methodologies and results
of the extensive simulation sets that have been done using Matlab.
We consider layered video composed with $L$ layers. The physical
overhead of IEEE 802.11a is illustrated in Table \ref{cap:IEEE 802.11a PHYparameters}.
The data rate is $6Mb/s$ and the control rate is $6Mb/s$. The EDCA
parameters of each $AC$ are presented in Table \ref{cap:Used-EDCA-parameters}.
Poisson distributed traffic consisting of 1024-bytes packets was generated
to each $AC$ regarding the selected mapping strategy.

\begin{table}[htbp]
\caption{\label{cap:IEEE 802.11a PHYparameters}IEEE 802.11a PHY/MAC parameters
used in simulation}
\centering{}{\scriptsize }%
\begin{tabular}{c|c|c|c}
\hline 
{\scriptsize SIFS} & {\scriptsize 16$\mu s$} & \textbf{\scriptsize $T_{PYS}$} & {\scriptsize 4$\mu s$}\tabularnewline
\hline 
{\scriptsize ACK size} & {\scriptsize 14 bytes} & \textbf{\scriptsize $T_{SYM}$} & {\scriptsize 4$\mu s$}\tabularnewline
\hline 
{\scriptsize PHY rate} & {\scriptsize 6 Mbits/s} & \textbf{\scriptsize $T_{P}$} & {\scriptsize 16$\mu s$}\tabularnewline
\hline 
{\scriptsize Slot-time} & {\scriptsize 9 $\mu s$} & \textbf{\scriptsize $\delta$} & {\scriptsize 1$\mu s$}\tabularnewline
\hline 
\end{tabular}
\end{table}

\begin{table}[b]
\caption{\label{cap:Used-EDCA-parameters}MAC parameters for the EDCA TCs.}

\centering{}{\scriptsize }%
\begin{tabular}{c|c|c|c|c}
\hline 
\textbf{\scriptsize Parameters/ACi} & \textbf{\scriptsize 0} & \textbf{\scriptsize 1} & \textbf{\scriptsize 2 } & \textbf{\scriptsize 3}\tabularnewline
\hline 
\textbf{\scriptsize $CW_{min}$} & {\scriptsize 7} & {\scriptsize 15} & {\scriptsize 31} & {\scriptsize 31}\tabularnewline
\hline 
\textbf{\scriptsize $CW_{max}$} & {\scriptsize 15} & {\scriptsize 31} & {\scriptsize 1023} & {\scriptsize 1023}\tabularnewline
\hline 
\textbf{\scriptsize AIFS{[}0,1,2,3{]}$(\mu s)$} & {\scriptsize 2} & {\scriptsize 2} & {\scriptsize 3} & {\scriptsize 7}\tabularnewline
\hline 
\textbf{\scriptsize Max-retry limit{[}0,1,2,3{]}} & {\scriptsize 7} & {\scriptsize 7} & {\scriptsize 7} & {\scriptsize 4}\tabularnewline
\hline 
\end{tabular}
\end{table}
We aim to evaluate the performance of different mapping strategies
(cano:OCOM, non-cano:selects the best mapping from non-canonical vectors,
and the sub-optimal algorithm) for layered video to different access
categories. In order to identify the adequate scheme that matches
the best mapping strategy, we implemented an algorithm that defines
all exhaustive mapping techniques described previously. For each described
mapping, we compute the average useful layers and the dropping probability
as defined in our analytical model. We classify the obtained set of
mapping strategies, to canonical mapping and non-canonical mapping.
For each simulation setting, and for each mapping group, we select
the best strategy minimizing the packets drop and maximizing the average
useful layers delivered. 

\begin{figure}[htbp]
\begin{centering}
\includegraphics[width=8cm,height=5cm,angle=360]{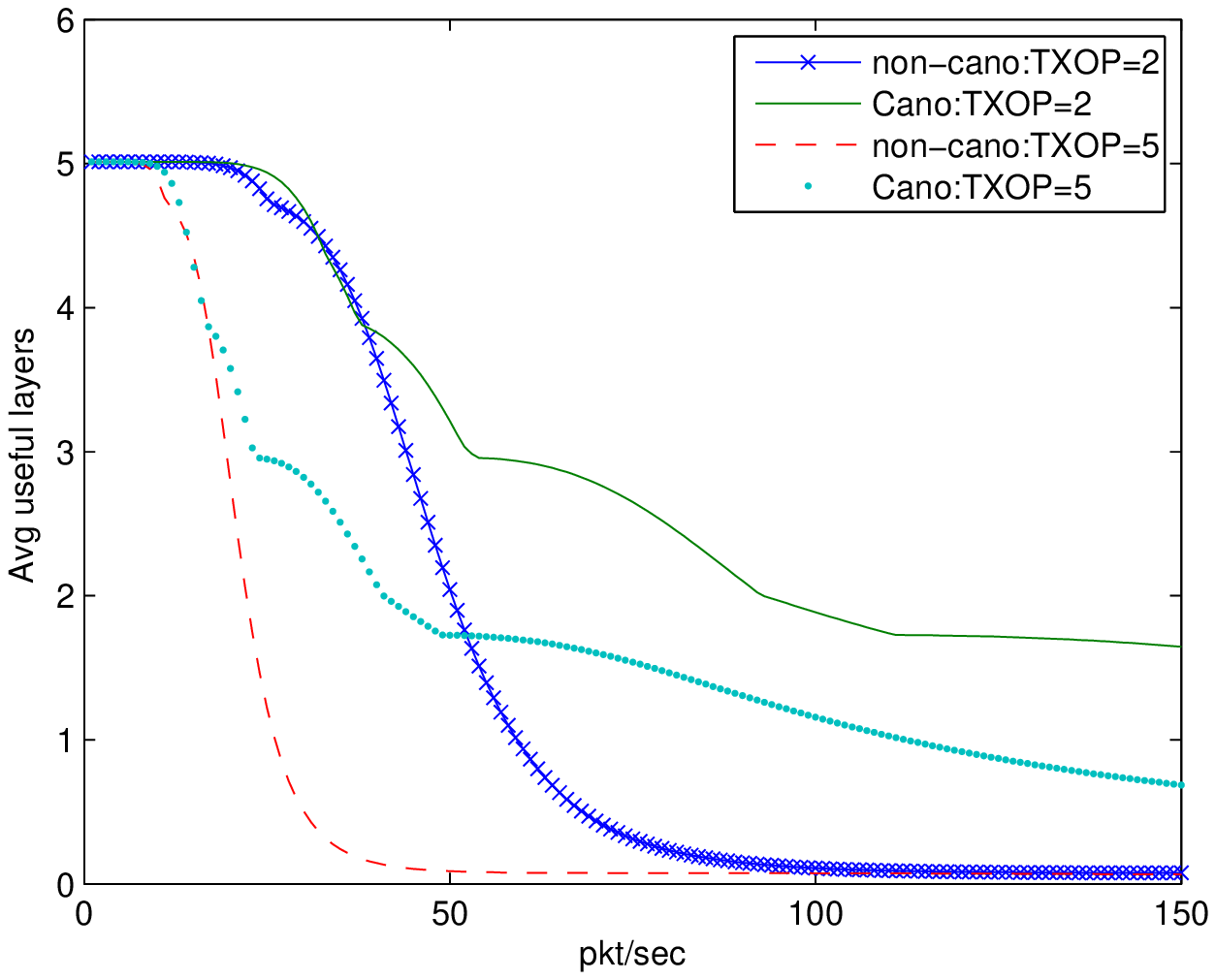}
\par\end{centering}

\caption{\label{cap:usl-10n_8vl}Average useful layers for N=10, L=8.}
\end{figure}

\begin{figure}[htbp]
\begin{centering}
\includegraphics[width=8cm,height=5cm,angle=360]{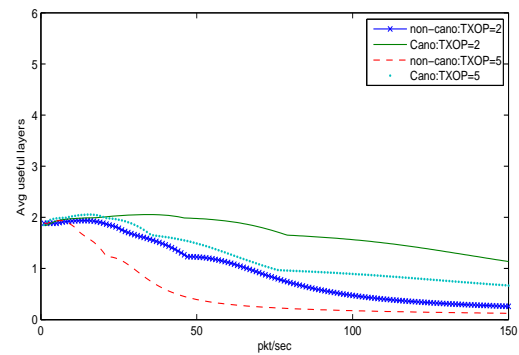}
\par\end{centering}

\caption{\label{cap:avg_ul-18nodes-8vl}Average useful layers N=18, L=8}
\end{figure}
\begin{figure}[htbp]
\begin{centering}
\includegraphics[width=8cm,height=5cm,angle=360]{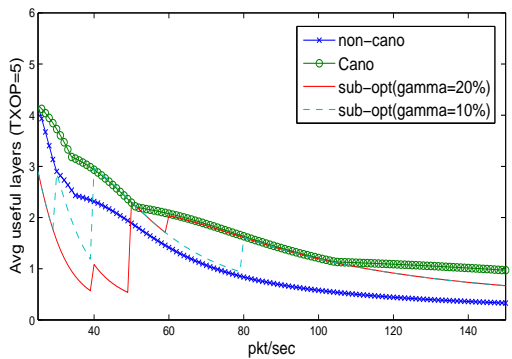}
\par\end{centering}

\caption{\label{cap:usl-5n_6vl-5txop-1}Average useful layers for N=5, L=6
.}
\end{figure}

Estimated number of useful layers successfully received by the destination,
is a good metric that informs strongly about video quality. We use
this metric to evaluate our proposals. We report results for $N=10$
and $N=18$ with various video application rates. We observe the evaluation
results in the case of video coding using 8 and 20 video layers, and
we consider different TXOP durations. The obtained results confirms
our analytical study, they show that the canonical mapping ensures
the best expected number of useful layers successfully delivered to
the destination.
\begin{figure}[htbp]
\begin{centering}
\includegraphics[width=8cm,height=5cm,angle=360]{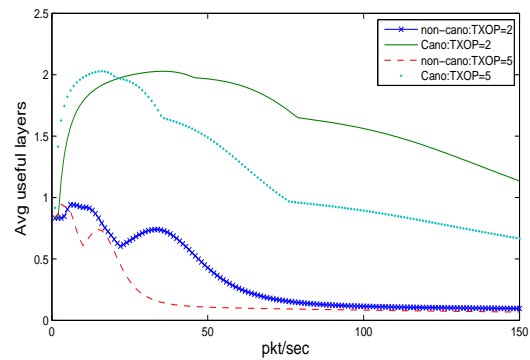}
\par\end{centering}

\caption{\label{cap:usl-18n_20vl-1}Average useful layers for N=18, L=20.}
\end{figure}
Moreover, Figure \ref{cap:avg_ul-18nodes-8vl} and Figure \ref{cap:usl-10n_8vl},
show that for low application rate with $TXOP=5$ and $TXOP=20$ the
results obtained using canonical and non-canonical mapping have the
same trends. Furthermore, the best video quality is obtained with
the lowest number of layers ($L=8$). Indeed, for low data rate, with
$8$ layers, the average useful layers is about $5$. However, when
considering $20$ layers, only about $2$ useful layers are delivered
successfully to the destination. A significant improvement is obtained
with canonical mapping when we increase the data rate, and using $TXOP=2$
gives better performance than using $TXOP=5$. Moreover, Figure \ref{cap:usl-18n_20vl-1}
shows that increasing the number of layers for the same number on
contending nodes does not enhance video quality as the average useful
layers is almost similar for both scenarios.
\begin{figure}[htbp]
\begin{centering}
\includegraphics[width=8cm,height=5cm,angle=360]{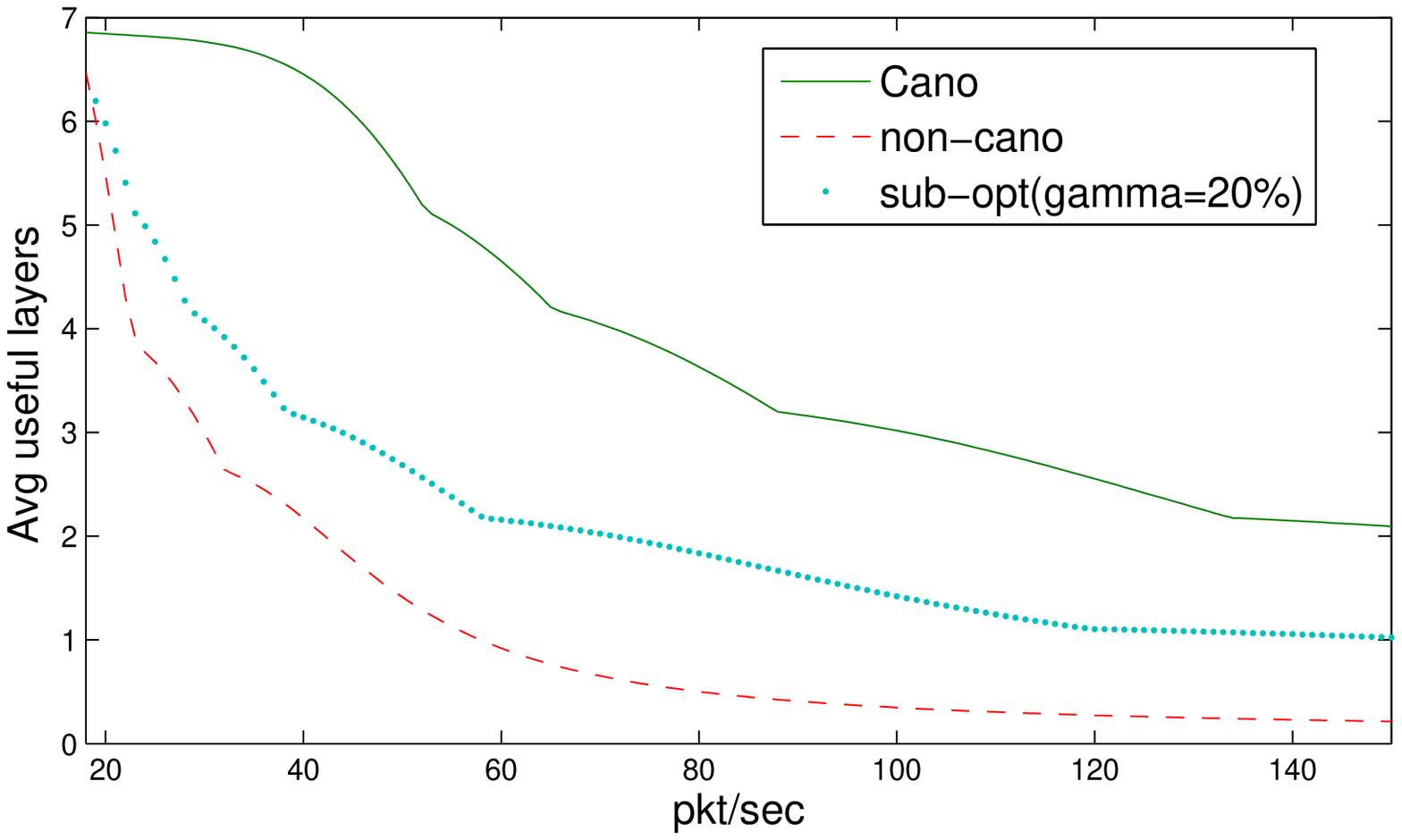}
\par\end{centering}

\caption{\label{cap:usl-10n_8vl-2txop}Average useful layers for N=10, L=8.}
\end{figure}
Figure \ref{cap:usl-10n_8vl-2txop} shows the results obtained using
canonical, non-canonical, and the sub-optimal algorithm described
above for $L=8$ and $N=10$. The sub-optimal proposal outperforms
the non-canonical mapping. The best performance results are obtained
with the canonical mechanism. Hence, the sub-optimal algorithm results
ensure a compromise between complexity and performance enhancement.
We set the threshold of the sub-optimal respectively to $\delta=10\%$
and $\delta=20\%$. Figure \ref{cap:usl-5n_6vl-5txop-1} shows that
for low bitrate, the sub-optimal algorithm using $\delta=10\%$ gives
better expected average useful layers than $\delta=20\%$, and both
results are lower than other mapping strategies results. However,
the performance of sub-optimal scheme increases when the application
rate increases and becomes similar to the canonical mapping strategy
results . 
Based on the obtained results, we propose to dynamically map the video
layers to different EDCA ACs regarding the estimated performance metric
(expected number of useful layers). The results showed that canonical
mapping strategy is recommended for high data rate and high number
of layers used for video coding. As, our analytical results show that
the best solution belongs to canonical mapping strategies, thus, instead
of performing an exhaustive search over all possible mapping combinations,
to select the best mapping strategy, only canonical mapping strategies
will be considered. This will decrease the complexity of the proposed
adaptive algorithm and optimize the calculation performed to obtain
best performance metrics. Furthermore, we proposed a simple sub-optimal
mapping algorithm based on heuristic study to more reduce the complexity
of computation. The selected strategy considered to transmit video
layers, is automatically adapted in the available channel resources. 
\section{\label{sec:Conclusion-and-Future}Conclusion and Future work}

In this framework, we proposed a distributed layered video mapping
technique, over EDCA ACs. The proposed algorithm dynamically maps
video layers to EDCA\textquoteright{}s appropriate ACs. The optimal
mapping strategy was selected based on the estimated maximum average
useful layers delivered to the destination node. We showed that canonical
mapping strategies ensure the best performance comparing to other
different mapping possibilities, especially for high application data
rate. The obtained results showed that the described algorithm helps
in meeting the performance improvement and also in decreasing the
packet drops. 
The implementation of this algorithm in our Qatar University wireless
mesh network regarding the network resources, channel sensitivity
and other feedback information in protocol optimization could be the
future work. 
\section{Acknoledgement}
This work is supported by Qatar National Research Fund (QNRF) No. 08-374-2-144.

\end{document}